%% file: main.tex
\documentclass[
    reprint,
    superscriptaddress,
    amsmath,
    amssymb,
    aps,
]{revtex4-2}

\usepackage{graphicx}
\usepackage{dcolumn}
\usepackage{bm}
\usepackage{amsmath}
\usepackage{physics}
\usepackage{ulem}

\usepackage{hyperref}
\hypersetup{colorlinks=true, citecolor=blue, urlcolor=blue, linkcolor=blue}

\DeclareMathOperator*{\argmax}{arg\,max}

\begin{document}

\preprint{APS/123-QED}

\title{Scalable and Site-Specific Frequency Tuning of Two-Level System Defects in Superconducting Qubit Arrays}

\newcommand{\UCBQNL}{Quantum Nanoelectronics Laboratory, Department of Physics, University of California at Berkeley, Berkeley, CA 94720, USA}
\newcommand{\LBLCRD}{Computational Research Division, Lawrence Berkeley National Laboratory, Berkeley, California 94720, USA}

\author{Larry Chen}
\thanks{These authors contributed equally to this work.}
\affiliation{\UCBQNL}
\affiliation{\LBLCRD}
\email{larrychen@berkeley.edu}

\author{Kan-Heng Lee}
\thanks{These authors contributed equally to this work.}
\affiliation{\LBLCRD}

\author{Chuan-Hong Liu}
\affiliation{\LBLCRD}

\author{Brian Marinelli}
\affiliation{\UCBQNL}

\author{Ravi K. Naik}
\affiliation{\LBLCRD}

\author{Ziqi Kang}
\affiliation{\UCBQNL}

\author{Noah Goss}
\affiliation{\UCBQNL}

\author{Hyunseong Kim}
\affiliation{\UCBQNL}
\affiliation{\LBLCRD}

\author{David I. Santiago}
\affiliation{\LBLCRD}

\author{Irfan Siddiqi}
\affiliation{\UCBQNL}
\affiliation{\LBLCRD}

\date{\today}

\begin{abstract}
\input{sections/abstract}
\end{abstract}

\maketitle

\section{Introduction}
\input{sections/M00_intro}

\section{Voltage-Biased TLS Defects}

\begin{figure}
    \centering
    \includegraphics[width=\columnwidth]{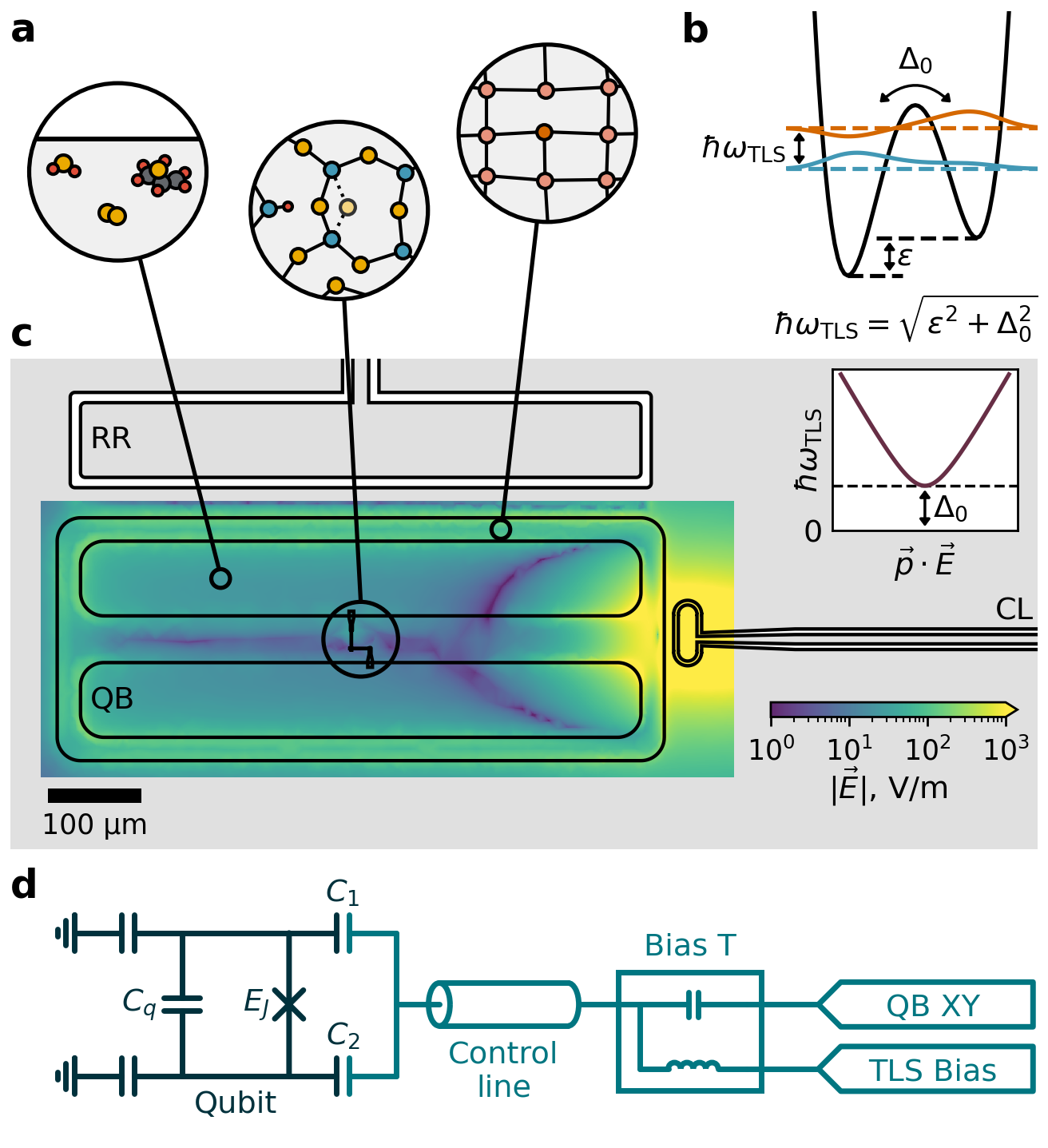}
    \caption{\textbf{Two-level systems in superconducting qubit devices.} \textbf{a,} Illustrations of proposed physical mechanisms for TLS in superconducting qubit devices, including (left to right): adsorbed molecules on the surface \cite{Kumar2016}, defects in the silicon substrate \cite{Zhang2024}, and defects in amorphous oxides \cite{Lisenfeld2023, Bal2024}. \textbf{b,} The double well potential for a TLS under the Standard Tunneling Model \cite{Anderson1972, Phillips1987, Müller2019}. The potential is characterized by an asymmetry energy $\varepsilon$, and a tunneling energy $\Delta_0$ that depends on the barrier height. \textbf{c,} A rendering of a single qubit on the experimental device showing the qubit capacitor pads (QB), the readout resonator coupling pad (RR), and the TIC-TAQ control line (CL). The simulated electric field distribution on the surface of the device is shown for a 1 V DC bias applied at the control line. The inset shows the dependence of the TLS frequency on the applied electric field, $\vec{E}$, and the TLS electric dipole moment, $\vec{p}$. \textbf{d,} An effective circuit diagram for the qubit/control-line system. The DC voltage bias for TLS control is combined with the RF signal for qubit XY control with an off-chip bias tee.
    }
    \label{fig:1}
\end{figure}

\input{sections/M01_tuning}

\begin{figure}
    \centering
    \includegraphics[width=\columnwidth]{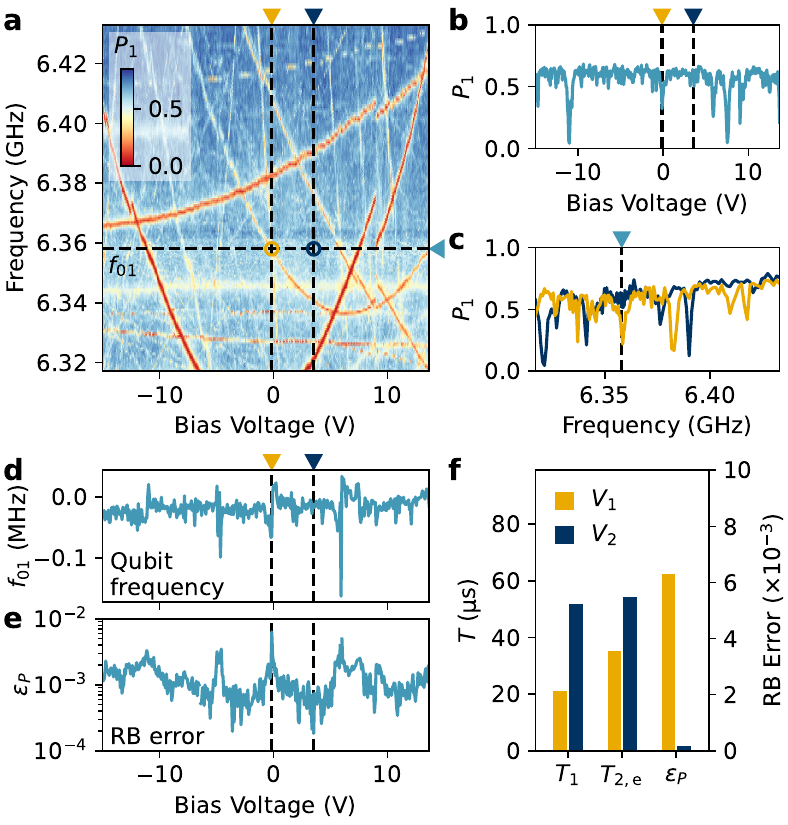}
    \caption{\textbf{Characterization of a qubit and its TLS environment under a varying voltage bias.} \textbf{a,} $T_1$ Stark shift spectroscopy as a function of applied voltage. The qubit $\ket{1}$-state population $P_1$ is measured at a fixed time of 20 \textmu s. The horizontal dashed line denotes the fundamental frequency, $f_{01}$, of the transmon qubit with no Stark shift. The vertical dashed lines mark two example voltage biases: $V_1 = -0.15$ V (yellow triangle) and $V_2 = 3.525$ V (dark blue triangle). Each acquisition of $P_1$ as a function of qubit frequency takes approximately 1 minute for a total acquisition time of approximately 11 hours. At $V_b = 9$ V, we see an abrupt change in the TLS spectrum, which may be due to a charge scrambling event \cite{Thorbeck2023}. \textbf{b,} Linecut of $P_1$ as a function of the bias voltage. At $V_1$ (yellow), we find a dip in $P_1$ corresponding to a TLS that is resonant with the qubit. In contrast, there is no resonant TLS at $V_2$ (dark blue). \textbf{c,} Linecuts of $P_1$ as a function of frequency at $V_1$ (yellow) and $V_2$ (dark blue). \textbf{d,} The qubit frequency, $f_{01}$, measured via ramsey interferometry as a function of the bias voltage. \textbf{e,} Single qubit Clifford randomized benchmarking (RB) as a function of the bias voltage. \textbf{f,} Qubit $T_1$, $T_{2, \text{echo}}$, and RB error rates ($\epsilon_P$) measured at $V_1=-0.15$ V (yellow) and $V_2 = 3.525$ V (dark blue).}
    \label{fig:2}
\end{figure}

\section{Temporal Stability}
\begin{figure}[ht]
    \centering
    \includegraphics[width=\columnwidth]{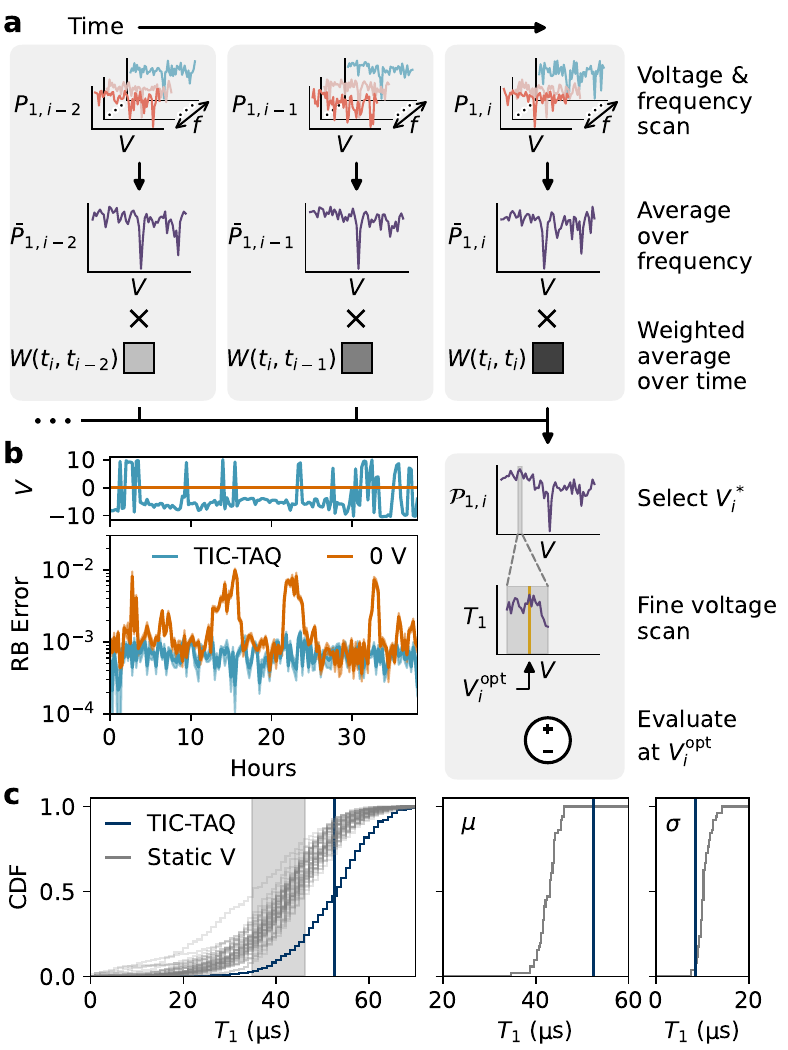}
    \caption{\textbf{Stabilizing qubit performance with TIC-TAQ.} \textbf{a,} A diagram illustrating the optimization procedure applied on qubit B1. \textbf{b,} The optimal voltages $V_i^\text{opt}$ (blue) as determined by the optimization algorithm as a function of time over a period of $\sim$38 hours (top panel). The static baseline voltage, $V_b = 0$, (orange) is shown as a reference. The RB error evaluated at $V_i^\text{opt}$ (blue) and $V_b = 0$ (orange) are shown in the bottom panel. \textbf{c,} Cumulative distributions of 500 repeated $T_1$ measurements taken over a 22 hour time period. The gray traces show the distributions measured at 51 static voltages, while the dark blue trace shows the distribution measured at the optimal voltage $V_i^\text{opt}$ for each iteration. The shaded gray region spans the minimum and maximum mean $T_1$ time for all static voltage configurations, and the vertical blue line denotes the mean $T_1$ time for the dynamic voltage configuration. The cumulative distributions of the means ($\mu$) and standard deviations ($\sigma$) for the static voltage biases are also plotted. The vertical line denotes the mean (standard deviation) for the dynamic voltage configuration.}
    \label{fig:3}
\end{figure}
\label{sec:stability}
\input{sections/M02_stability}

\begin{figure*}[ht]
    \centering
    \includegraphics[width=\textwidth]{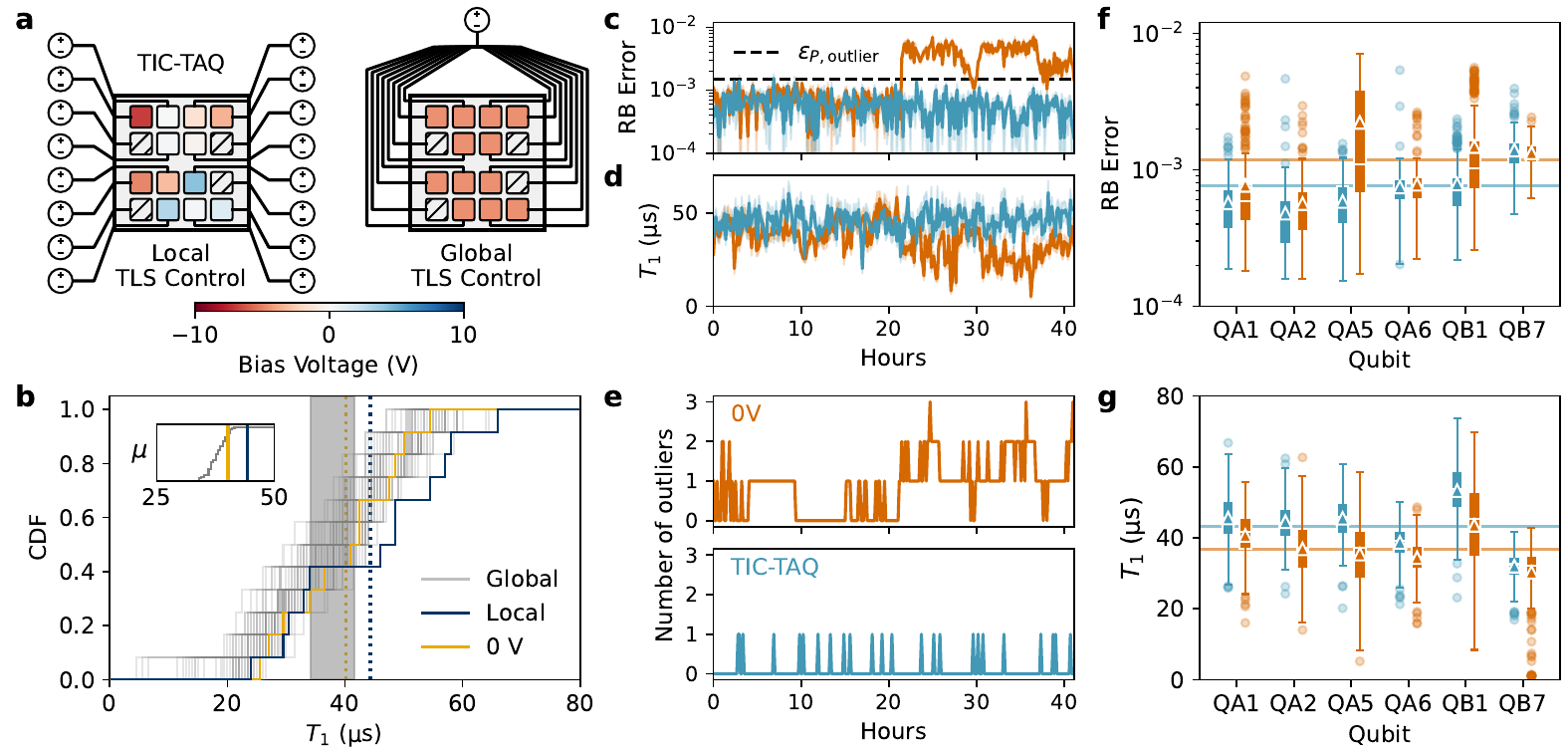}
    \caption{\textbf{Benchmarking TIC-TAQ across a multi-qubit device.} \textbf{a,} A schematic illustrating the locally optimized (left) and global (right) voltage bias configurations. The 4 hatched squares represent qubits that were not measured (see Appendix Sec. \ref{sec:device} for details). \textbf{b,} A comparison of the cumulative distributions of $T_1$ times across 12 qubits with local optimization (dark blue) compared to global optimization (gray). The yellow trace highlights the global configuration where $V_b = 0$ for all qubits. The dotted lines denote the means of the corresponding distributions, and the grey shaded region spans the minimum and maximum mean $T_1$ times for all global configurations. The inset shows the distribution of mean $T_1$ times across the chip, with the mean for the $V_b = 0$ V (yellow) and locally optimized (dark blue) configurations marked by vertical lines. \textbf{c,d,} Repeated measurements of single qubit gate error ($T_1$) on qubit A5, measured at $V_b = 0$ V (orange) and $V_i^\text{opt}$ (blue). The black dashed line in \textbf{c} denotes the outlier threshold for this qubit, defined in Sec. \ref{sec:stability}. \textbf{e,} The number of outlier qubits across the 6 qubit subset over 40 hours. \textbf{f,g,} The distributions of single qubit gate errors ($T_1$ times) for each qubit. The circles denote outliers in the distribution and the horizontal lines denote the mean RB error ($T_1$ times) across all 6 qubits.}
    \label{fig:4}
\end{figure*}

\section{Device-Wide Performance}
\input{sections/M03_scaling}

\section{Discussion}
\input{sections/M04_discussion}

\section*{Author Contributions}
L. Chen conceived the experiment, designed the device, and performed the measurements.
K. H. Lee and Z. Kang fabricated the experimental device with assistance from L. Chen.
L. Chen analyzed the results with help from B. Marinelli, R. K. Naik, C. H. Liu and K. H. Lee. L. Chen and B. Marinelli derived the circuit analysis for the control line.
K. H. Lee developed the finite element simulations of the control line designs.
C. H. Liu developed TLS fitting routine.
L. Chen, K. H. Lee, C. H. Liu, B. Marinelli and R. K. Naik wrote the manuscript with input from all authors.
D. I. Santiago and I. Siddiqi supervised the experimental effort.
All authors contributed to the writing and editing of the manuscript.

\begin{acknowledgments}
The authors would like to thank Alexis Morvan, Gerwin Koolstra, and Long Nguyen for valuable feedback on the manuscript. We thank Bingcheng Qing for helpful discussions regarding device fabrication, and Thomas Ersevim for assistance with the experimental setup. This material was funded by the U.S. Department of Energy, Office of Science, Office of Advanced Scientific Computing Research Quantum Testbed Program under contract DE-AC02-05CH11231.
\end{acknowledgments}

\bibliography{bib}

\clearpage
\appendix

\setcounter{figure}{0}
\renewcommand{\thefigure}{S\arabic{figure}}

\section{Device and Fabrication Details}
\begin{figure}[ht]
    \centering
    \includegraphics[width=\columnwidth]{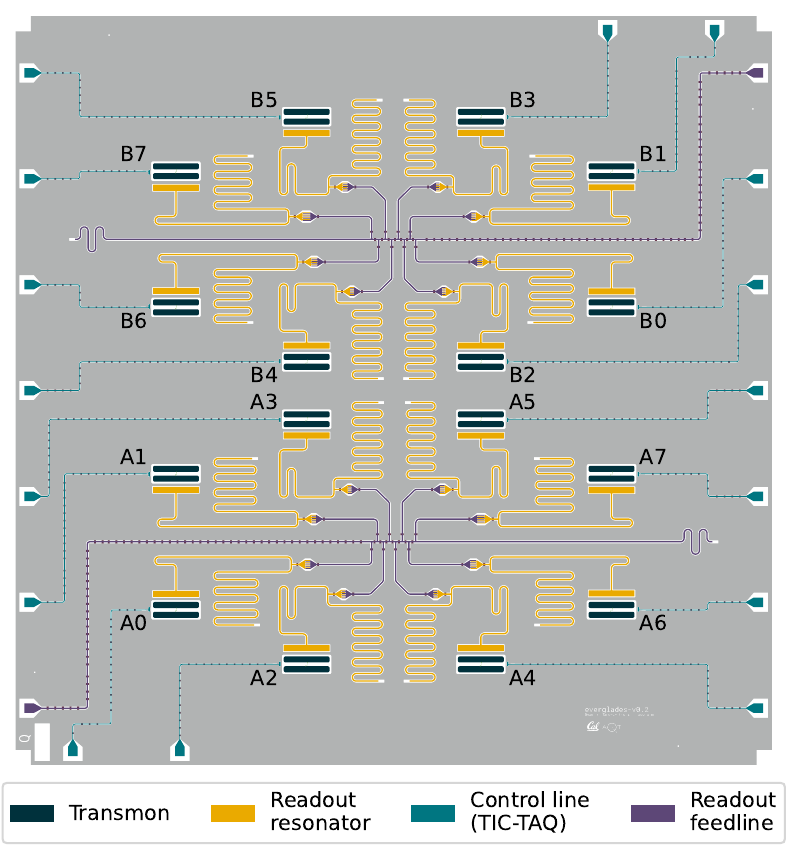}
    \caption{\textbf{Device schematic.} A schematic of the 16-qubit device used in this work. Qubits A0, A7, B0, and B6 did not survive the fabrication process and could not be measured.}
    \label{fig:chip}
\end{figure}
\input{sections/S01_device}

\section{Experimental Setup}
\begin{figure}[ht]
    \centering
    \includegraphics[width=\columnwidth]{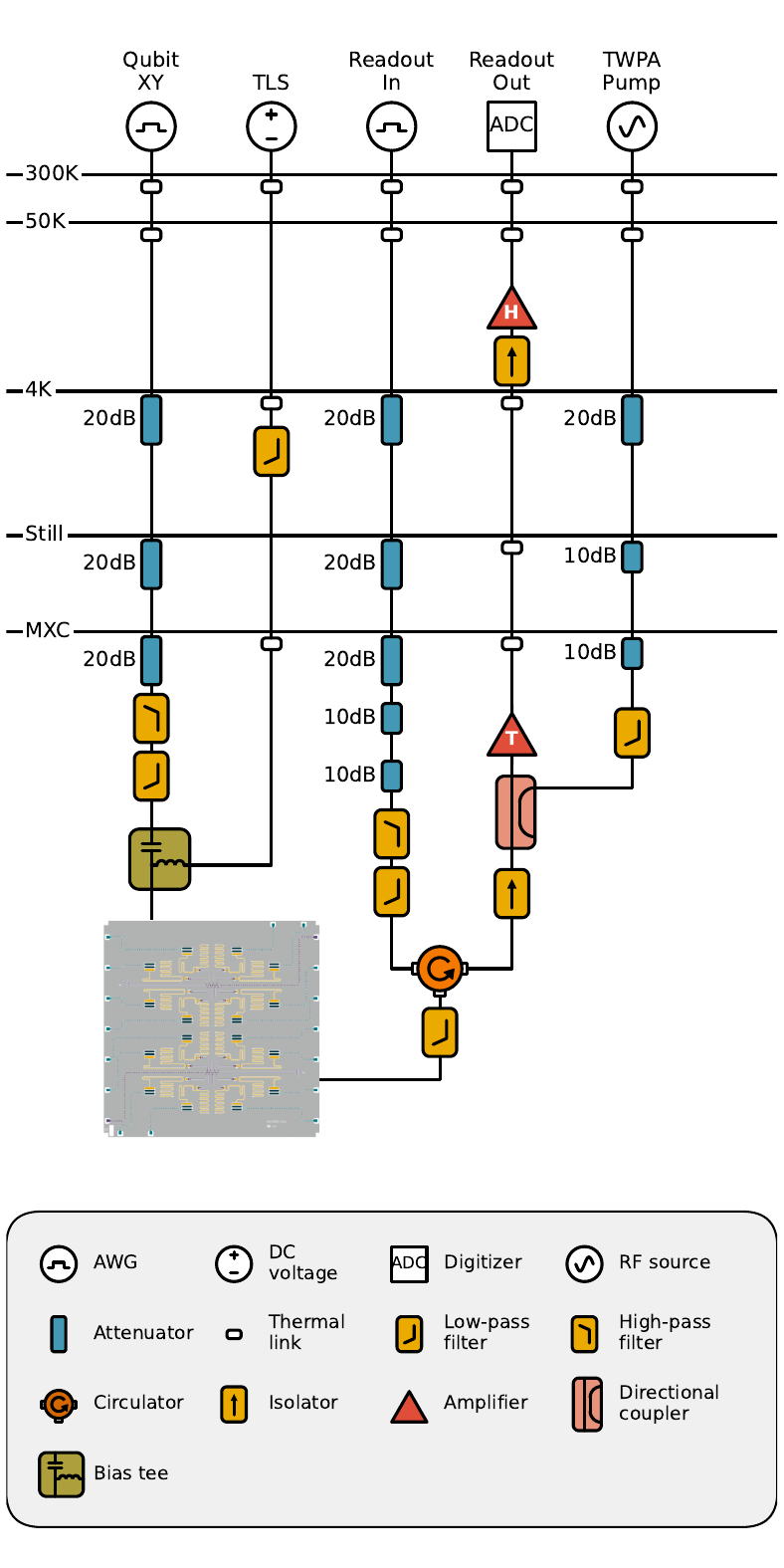}
    \caption{\textbf{Wiring diagram.} A schematic detailing the wiring, filters, and control components used for this experiment.}
    \label{fig:chip}
\end{figure}
\input{sections/S02_experimental-setup}

\section{Multiplexed Control Line Design}
\input{sections/S03_control-line}

\begin{figure}[ht]
    \centering
    \includegraphics[width=\columnwidth]{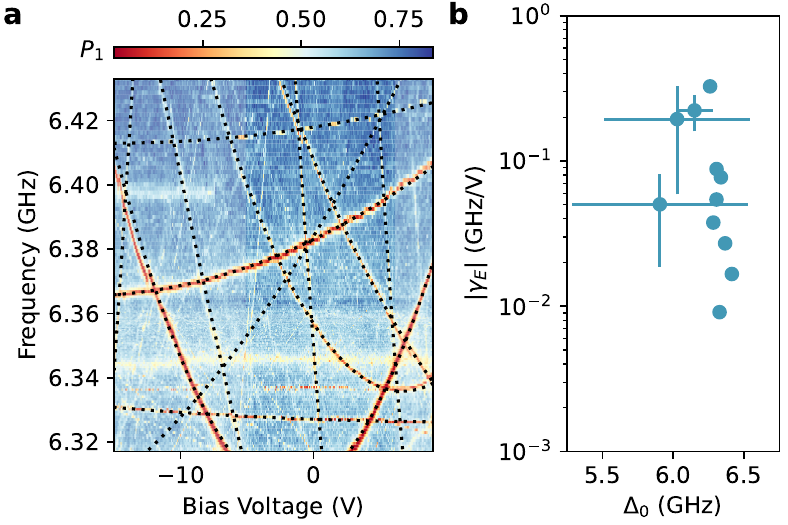}
    \caption{\textbf{Fits of TLS tuning curves} \textbf{a,} An example TLS spectrum as a function of the applied voltage, from the same dataset shown in Fig. \ref{fig:2}\textbf{a}. The black dotted lines indicate fits to the STM (Eq. \ref{eq:stm}). We exclude the data between 9 V and 13.65 V from the fit, where the TLS spectrum exhibits a discontinuous shift. \textbf{b,} A scatter plot showing the extracted fit parameters $\gamma_E$ (tuning sensitivity) and $\Delta_0$ (tunneling energy) for 11 defects. Error bars indicate fit uncertainties.}
    \label{fig:tls-fit}
\end{figure}

\section{TLS Spectrum Fitting}
\input{sections/S04_TLS-fitting}

\section{DC Crosstalk}
\begin{figure*}[ht]
    \centering
    \includegraphics[width=\textwidth]{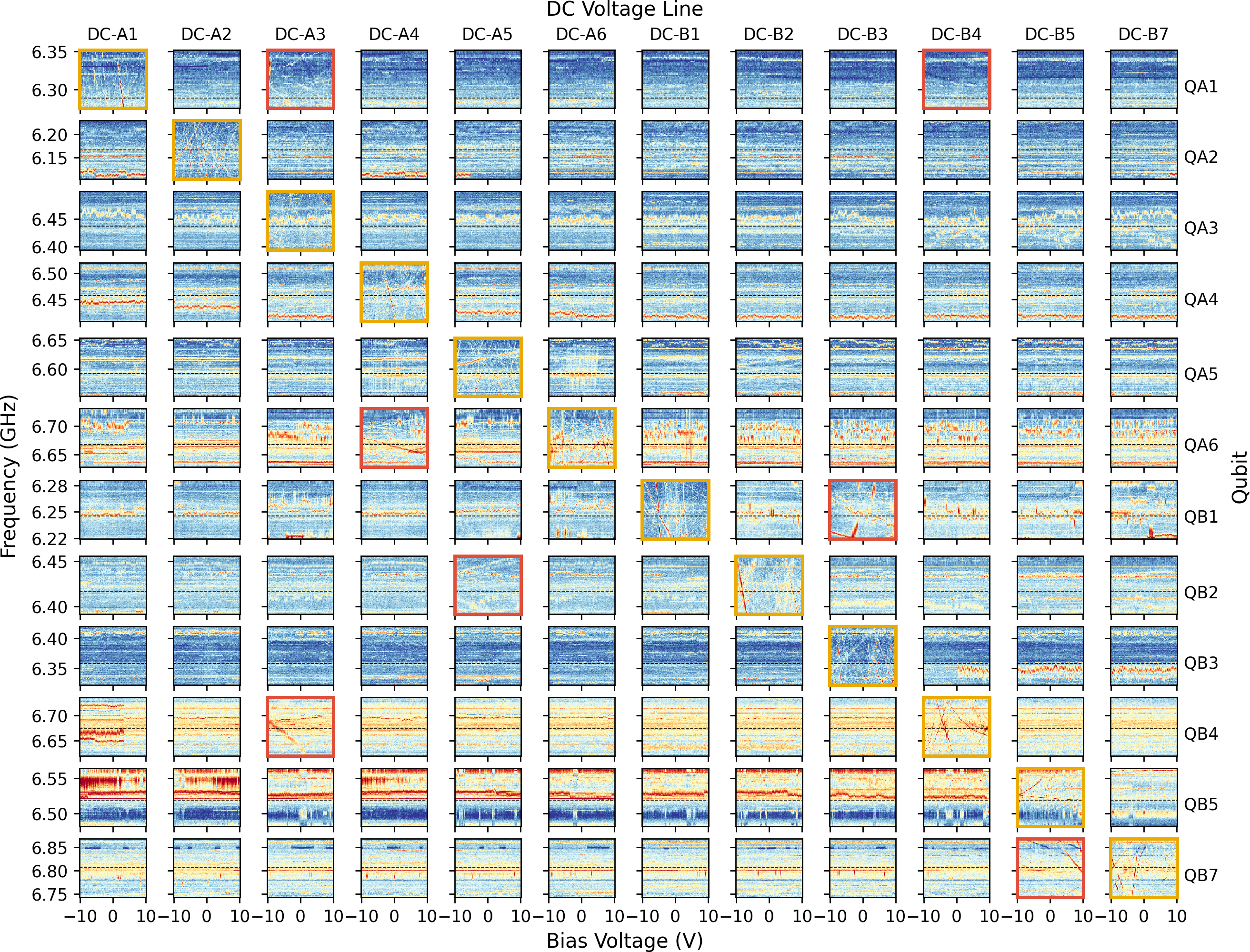}
    \caption{\textbf{DC Crosstalk Matrix.} TLS spectroscopy data showing the TLS spectrum on each qubit as a function of each control line. The yellow borders along the diagonal indicate the intended control-line/qubit pairs. We highlight notable control-line/qubit pairs that exhibit unwanted tuning of the TLS spectrum with a red border. The data in each column was acquired concurrently.}
    \label{fig:crosstalk}
\end{figure*}
\input{sections/S05_crosstalk}

\section{Optimization Details}
\input{sections/S06_optimization}

\end{document}

%% file: sections/abstract.tex
State-of-the-art superconducting quantum processors containing tens to hundreds of qubits have demonstrated the building blocks for realizing fault-tolerant quantum computation. Nonetheless, a fundamental barrier to scaling further is the prevalence of fluctuating quantum two-level system (TLS) defects that can couple resonantly to qubits, causing excess decoherence and enhanced gate errors. Here we introduce a scalable architecture for site-specific and in-situ manipulation of TLS frequencies out of the spectral vicinity of our qubits. Our method is resource efficient, combining TLS frequency tuning and universal single qubit control into a single on-chip control line per qubit. We independently control each qubit's dissipative environment to dynamically improve both qubit coherence times and single qubit gate fidelities --- with a constant time overhead that does not scale with the device size. Over a period of 40 hours across 6 qubits, we demonstrate a $36\%$ improvement in average single qubit error rates and a $17\%$ improvement in average energy relaxation times. Critically, we realize a 4-fold suppression in the occurrence of TLS-induced performance outliers, and a complete reduction of simultaneous outlier events. These results mark a significant step toward overcoming the challenges that TLS defects pose to scaling superconducting quantum processors.

%% file: sections/M00_intro.tex
Over the last two decades, remarkable progress has been made toward engineering high-coherence superconducting qubit devices at scale \cite{Siddiqi2021, Devoret2013}. These advances have enabled demonstrations of quantum advantage \cite{Google2019, Zhu2022, Morvan2024}, implementations of error-corrected quantum memories beyond break-even \cite{Google2024}, and simulations of quantum many-body systems \cite{Kim2023}. However, noise continues to limit both device performance and stability, limiting potential demonstrations of quantum utility.

A dominant source of noise for superconducting qubits stems from interactions with a large bath of two-level system (TLS) defects \cite{Martinis2005, Barends2013, Müller2019}. These microscopic defects exist primarily within the surfaces and interfaces of materials composing the qubits, where they can interact strongly with the qubit's electromagnetic field \cite{Gao2008, Wang2009}. When a TLS resonance is near the qubit frequency, it can lead to enhanced decoherence \cite{Barends2013}, a common source of performance outliers, or \textit{dropouts}, in a multi-qubit processor \cite{Klimov2024, Mohseni2024}. The number of dropouts scales with the system size \cite{Mohseni2024}, posing a challenge for quantum error correcting codes by reducing the effective connectivity of a qubit lattice \cite{Debroy2024, Auger2017}. Furthermore, the noise imparted on qubits from defects is temporally unstable \cite{Klimov2018, Burnett2019, Carroll2022, Thorbeck2023, Zanuz2024}, requiring frequent recalibration to suppress outliers \cite{Klimov2024}. Reliably mitigating the impact of noise resulting from TLS defects is therefore essential for scaling superconducting circuits to larger system sizes.

Two complementary approaches have emerged in the effort to address this challenge. The first focuses on reducing the average density of TLS defects and their coupling to the qubit. This includes materials engineering to develop higher-quality fabrication processes \cite{Place2021, Altoé2022, Crowley2023, Odeh2025} and design improvements to reduce qubit sensitivity to dissipative interfaces \cite{Wenner2011, Wang2015, Dial2016}. While these efforts have successfully improved average coherence times, spatial and temporal outliers induced by strongly-coupled TLSs continue to plague state-of-the-art quantum processors \cite{Mohseni2024}.

The second approach consists of \textit{in-situ} control strategies to actively suppress the impact of excess TLS noise during qubit operation. For example, the operating frequencies of individual qubits can be optimized to avoid collisions with relaxation hot-spots caused by TLS defects \cite{Klimov2020, Zhao2022, Klimov2024}. However, this is a highly-constrained problem with non-local scope, because the optimal operating frequencies for individual qubits must also account for other frequency-dependent error sources \cite{Klimov2024}. Moreover, updates to the qubit frequency require additional calibration or modeling overhead to minimize coherent control errors, since qubit control parameters also depend on the operating frequency \cite{Rol2020, Hyyppä2024}. Alternatively, several studies have demonstrated the spectral manipulation of TLS defects using global electric fields \cite{Lisenfeld2019, Bilmes2020, Bilmes2021, Bilmes2022}, global mechanical strain \cite{Lisenfeld2016, Lisenfeld2019, Bilmes2021, Bilmes2022}, or other unspecified control parameters \cite{Kim2024}. However, since the TLS spectral landscape varies from qubit to qubit, methods for globally manipulating the TLS spectra of many qubits with a single control parameter are unlikely to succeed. Therefore, independent and local control over each qubit's noise environment will be critical for addressing the TLS problem at scale.

Here we introduce a scalable architecture for the \textit{targeted in-situ control of TLS and qubits} (TIC-TAQ), enabling simultaneous and independent control of each qubit and its TLS noise environment through a single on-chip control line. We use this novel control architecture to apply a site-specific electric field that can detune TLSs from the qubit transition frequency and mitigate TLS-induced noise. By dynamically manipulating the TLS spectrum, we demonstrate robustness against excess decoherence and show reduced fluctuations in both qubit lifetimes and single-qubit gate errors. We use this control architecture to demonstrate the simultaneous suppression of TLS-induced errors across 6 qubits in a multi-qubit device, over a time period of 40 hours. Our approach offers a simple yet effective tool for mitigating the damaging effects of TLS defects in a quantum processor.

%% file: sections/M01_tuning.tex
\label{sec:tuning}

While the precise microscopic origins of TLSs in superconducting qubits are unknown and many possibilities have been suggested (Fig. \ref{fig:1}\textbf{a}), the universal behavior of these defects in amorphous materials can be described by the Standard Tunneling Model \cite{Anderson1972, Phillips1987, Müller2019}, which treats the TLS as a quantum system that can tunnel between two potential wells (Fig. \ref{fig:1}\textbf{b}). Here, the frequency of the TLS is 
\begin{equation}
    \hbar\omega_\text{TLS} = \sqrt{\varepsilon^2 + \Delta_0^2},
\end{equation}
where $\Delta_0$ is the tunneling energy and $\varepsilon$ defines the asymmetry between the wells. The asymmetry energy of the TLS is modified by its coupling to an external strain field $\mathbf{S}$ or electric field $\mathbf{E}$ as
\begin{equation}
\label{eq:stm}
    \varepsilon = 2\mathbf{\gamma}\cdot\mathbf{S} + 2\mathbf{p}\cdot\mathbf{E} + \varepsilon_0,
\end{equation}
where $\mathbf{\gamma}$ describes the coupling of the TLS to the strain field, $\mathbf{p}$ is the electric dipole moment of the TLS, and $\varepsilon_0$ is an offset that depends on the TLS's local environment. As a result, TLSs exhibit a hyperbolic frequency dependence when exposed to an external electric field (Fig. \ref{fig:1}\textbf{c}).

With the TIC-TAQ architecture, we can realize strong local electric fields that couple to the qubit's TLS environment, while preserving qubit coherence, by exploiting the symmetry of the quantum circuit. Here, quantum information is encoded in the differential mode of a floating transmon circuit, which consists of a Josesphon junction shunted by the capacitance between two superconducting electrodes (Fig. \ref{fig:1}\textbf{c}). Fig. \ref{fig:1}d illustrates the circuit diagram for the transmon-control line system. The effective coupling between the control line and the differential qubit mode is proportional to the difference, $C_1 - C_2$, of the capacitances to each pad, while the coupling to TLSs on the surfaces of the qubit electrodes is proportional to the sum, $C_1 + C_2$. We leverage the symmetry in the qubit structure to increase the coupling capacitance between the control line and each electrode to $\sim$0.5 fF, more than an order of magnitude larger than a typical coupling capacitance of ~25 aF for a grounded transmon \cite{Bardin2020}, while maintaining Purcell limited $T_1$ times well above 1 ms (see Appendix Sec. \ref{sec:cl-design} for more details on the design). With this design, a 1 V voltage bias applied on the control line can induce electric fields $|\mathbf{E}| > 130$ V/m even at the far edge of the qubit pad, according to finite element simulations (Fig. \ref{fig:1}\textbf{c}). This corresponds to a tuning sensitivity, $\gamma_E \sim 4$ MHz/V for a typical dipole moment $p_\text{TLS} \sim 3$ debye \cite{Sarabi2016, Brehm2017}.

We study the effect of applying a DC voltage bias on the control line with AC Stark shift spectroscopy \cite{Carroll2022}, a variation of swap spectroscopy \cite{Barends2013} that harnesses the off-resonant interaction between a microwave drive and a qubit to shift its frequency. This technique allows us to probe the spectrally resolved lifetimes of the transmon 0-1 transition over a $\sim$100 MHz range and identify individual defects that show up as dips in the $T_1$ spectrum. Using the transmon as a TLS sensor, we sweep the applied voltage bias on the control line over a 30 V range. Fig. \ref{fig:2}\textbf{a} shows the transition frequencies of TLS defects coupled to an example transmon as a function of the applied voltage bias. We observe a range of tuning sensitivities $\gamma_E$ between $9$ MHz/V and $330$ MHz/V (see Appendix Sec. \ref{fig:tls-fit}). We show example linecuts at the qubit frequency (Fig. \ref{fig:2}\textbf{b}) and constant voltage (Fig. \ref{fig:2}\textbf{c}) to better illustrate the response of the qubit's dissipative environment to the applied voltage.

Next, we examine how the target qubit is affected by a DC voltage bias on its control line, which will also induce an offset charge, $n_g$, across the junction. However, the exponentially suppressed charge dispersion of the transmon ensures that the qubit is unaffected by the induced offset charge \cite{Koch2007}. In Fig. \ref{fig:2}\textbf{d}, we confirm that the qubit frequency, $f_{01}$ remains mostly unaffected by the applied voltage, apart from avoided level crossings caused by resonantly coupled TLSs. Repeated applications of single qubit Clifford randomized benchmarking (RB) \cite{Knill2008, Hashim2024} also show no direct dependence on the applied DC voltage (Fig. \ref{fig:2}\textbf{e}), suggesting that qubit control is not adversely affected by the DC electric field. Rather, the applied voltage bias primarily impacts the qubit by coupling to and modifying its dissipative environment.

The independent control of the qubit's noise environment with respect to its intrinsic properties allows for the unconstrained optimization of important performance metrics, including the qubit $T_1$ time, $T_{2,\text{e}}$ time, and single qubit gate fidelities. In Fig. \ref{fig:2}\textbf{f}, we compare these metrics at two example voltage biases. At $V_1 = -0.15$ V, where the qubit is approximately resonant with a TLS, we see a 60\% (35\%) reduction in the qubit's $T_1$ ($T_{2,\text{e}}$) time compared to $V_2 = 3.525$ V, where there is no resonant TLS. We also find a 35-fold increase in the single qubit RB error when the TLS is on-resonance, from both the reduced coherence times and coherent errors caused by the TLS-induced frequency shift. These TLS-induced errors typify dropout qubits, which limit quantum processor performance at scale.

%% file: sections/M02_stability.tex
Next, we use TIC-TAQ to address the problem of temporal stability for superconducting qubits. TLSs do not remain stationary, and fluctuations in each qubit's TLS environment are the primary driver of performance instability for superconducting circuits \cite{Klimov2018,Carroll2022,Burnett2019}. The TLS spectral landscape for a given qubit can change drastically on the timescale of hours, due to spectral diffusion \cite{Klimov2018} or charge scrambling events \cite{Thorbeck2023}. When these changes bring a strongly coupled TLS into resonance with a qubit for a prolonged period of time, it can render the qubit effectively inoperable.

In response to the dynamic TLS environment, we leverage the TIC-TAQ architecture to apply a variable electric field that biases TLS frequencies away from the qubit resonance, using the two-step optimization procedure illustrated in Fig. \ref{fig:3}\textbf{a}. 
For each time step, $t_i$, we first do a coarse voltage scan (-10 V to 10 V). At each bias voltage, we perform AC Stark shift spectroscopy to characterize the TLS landscape over a small frequency window ($f_{01} \pm 5$ MHz), which provides information about nearly resonant defects. We then compute an averaged signal $\bar{P}_{1,i}(V)$ over the sampled frequencies, which effectively smooths the signal over voltage. Next, we apply a weighted average over the $\bar{P}_1(V)$ from previous iterations to compute the loss function $\mathcal{P}_{1,i}(V)$ used to determine an initial estimate $V_i^*$ of the optimal voltage bias. This reduces statistical noise and accounts for historical information regarding the TLS landscape. Finally, we refine the initial estimate $V_i^*$ of the optimal bias voltage by measuring the qubit $T_1$ time in a small window around $V_i^*$ and selecting the voltage that yields the highest $T_1$ as $V_i^\text{opt}$. For our choice of hyperparameters (see Appendix Sec. \ref{sec:optimization}), the process of determining $V_i^\text{opt}$ takes approximately 200 seconds per iteration.

In Fig. \ref{fig:3}\textbf{b}, we show the optimized voltages $V_i^\text{opt}$ over a period of $\sim$38 hours, where we reoptimize approximately once every 15 minutes. Between iterations, we evaluate the single qubit gate errors at both $V_i^\text{opt}$ and a baseline voltage, $V_\text{b} = 0$ V. Over the full measurement period, we measure a mean gate error of $1.8\times 10^{-3}$ for the baseline configuration, which is 2.6$\times$ larger than the mean gate error of $6.8 \times 10^{-4}$ that we find for the optimized voltages. Moreover, at $V_b = 0$, we find many time periods with elevated single qubit gate errors, which we attribute to the presence of a TLS that intermittently moves into resonance with the qubit frequency. In contrast, the single qubit gate errors are relatively stable for the optimized voltages. 

We can quantify these temporal outliers by defining an outlier threshold, $\varepsilon_{P, \text{outlier}}$, based on the distribution of gate errors for the optimized configuration, $\varepsilon_{P,\text{outlier}} = Q_3 + 1.5 \times (Q_3 - Q_1)$,
where $Q_1$ ($Q_3$) is the first (third) quartile and $Q_3 - Q_1$ is the interquartile range (IQR). For the dataset shown in Fig. \ref{fig:3}\textbf{c}, we calculate an outlier threshold of $\varepsilon_{P, \text{outlier}} = 1.26 \times 10^{-3}$, detecting an outlier 36\% (1.3\%) of the time for the baseline (optimized) configuration.

To validate that the performance improvement demonstrated by the optimizer is not simply due to the particular choice of baseline voltage, we perform a separate experiment where we repeatedly evaluate the qubit $T_1$ time at 51 static voltage biases ($-10$ V to $10$ V) over a time period of $\sim$22 hours. Fig. \ref{fig:3}\textbf{c} shows the resulting $T_1$ distributions. We find a mean $T_1$ of 52.5 \textmu s at the optimized voltages, which represents a 14\% (38\%) improvement compared to the best (worst) mean $T_1$ across all static biases. Moreover, we find a standard deviation, $\sigma_\text{dynamic} = 8.4$ \textmu s, for the optimized voltages that is at the lower end of the distribution over static biases ($\sigma_\text{static} \in [7.7, 14.4]$ \textmu s) reflecting the relative stability enabled by dynamic mitigation. The different static bias conditions allow us to simulate the effect of independent cooldowns, which have been shown to change the precise frequencies of TLS defects but not their overall spectral densities \cite{Shalibo2010,Zanuz2024}. Therefore, the clear and consistent improvement in both the average and spread of $T_1$ times with dynamic mitigation confirms that the TIC-TAQ architecture can enable a reliable improvement in qubit performance.

%% file: sections/M03_scaling.tex
Having demonstrated the successful mitigation of TLS-induced errors on a single qubit, we now address the scalability of our technique to a large-scale quantum processor, whose overall performance can be strongly affected by a small group of poorly-performing outlier qubits \cite{Mohseni2024}. While there have been efforts to mitigate the harm caused by such dropouts in the context of quantum error correction \cite{Auger2017, Debroy2024}, performance inevitably worsens as their number increases. We show how TIC-TAQ can address this problem by simultaneously mitigating TLS-induced dropouts across multiple qubits with only constant calibration overhead.

In a multi-qubit device, each qubit couples to a separate and independent bath of TLS defects. Therefore, it is not expected that a single voltage bias is optimal for every qubit --- a voltage bias that tunes a TLS out of resonance with one qubit may tune a different TLS \textit{into} resonance with another qubit. This problem worsens as the TLS spectral density or the total number of qubits increases, suggesting that local optimization of each qubit's individual TLS environment is necessary to maximize the performance of a multi-qubit device.

We verify this with an experiment on 12 qubits comparing local TLS control, where each qubit's TLS environment is controlled with a site-specific voltage bias, and global TLS control, where all qubits share a common voltage bias (Fig. \ref{fig:4}\textbf{a}). To determine the optimal local bias for each qubit, we perform the optimization procedure described in Section \ref{sec:stability}, concurrently monitoring each qubit for $\sim$26 minutes. We use the selected voltages for each qubit from the final iteration to benchmark the local TLS control.

Fig. \ref{fig:4}\textbf{b} shows the distributions of average $T_1$ times (over $\sim$105 minutes) across 12 qubits. We compare the distributions of mean $T_1$ times across the device for 51 global bias values between $-10$ V and $10$ V (grey) to the distribution of average $T_1$ times for the locally optimized voltage biases (dark blue). We find a mean $T_1$ of 44.3 \textmu s across the 12 qubits for the local configuration, compared to a best (worst) mean $T_1$ of 41.6 \textmu s (34.1 \textmu s) for the global configurations, representing a $6.5\ -\ 30\%$ improvement in mean $T_1$ compared to all global configurations. This consistent improvement in mean $T_1$ across the device compared to all global configurations confirms the advantage of local and independent control over each qubit's individual TLS environment.

While the locally optimized configuration led to a better average $T_1$ across the entire device, it did not strictly improve the $T_1$ for each individual qubit. This can be explained in part by changing TLS landscapes, but is also the result of DC crosstalk between a subset of the qubit control lines (see Appendix Sec. \ref{sec:crosstalk}). Crosstalk breaks the assumption that each control line independently controls only a single qubit's TLS environment, which limits the performance of our parallel optimization method. However, this does not represent a fundamental limitation of our method, and we expect a straightforward suppression of unwanted crosstalk with the integration of superconducting through-silicon vias in future devices \cite{Hazard2023}.

As a final benchmark of the TIC-TAQ architecture, we repeat the experiment discussed in Section \ref{sec:stability} simultaneously on 6 qubits (selected to minimize crosstalk) for a total duration of $\sim$40 hours. Between each iteration, we measure the qubit $T_1$ and perform isolated single qubit RB on each qubit at both the periodically optimized voltages $V^\text{opt}_{i, q}$ (where $q$ represents the qubit index) and $V_b = 0$ V. Fig. \ref{fig:4}\textbf{c}(\textbf{d}) shows the time-series of RB errors ($T_1$ times) for a representative qubit (A5). For the static configuration, $V_b = 0$ V (orange), we find a sudden drop in the qubit $T_1$ and a corresponding spike in the single qubit gate error beginning at approximately $t = 21$ hours. In contrast, the qubit performance remains stable and unaffected by changes in the TLS spectrum when the voltage bias is periodically optimized (blue).

For each qubit, we calculate an outlier threshold (as defined in Sec. \ref{sec:stability}) to screen for TLS-induced outlier events. Fig. \ref{fig:4}\textbf{e} shows the number of outlier qubits over time for both voltage configurations. Over a period of $\sim$40 hours, we detect one or more outlier qubits only 8.2\% of the time with periodic optimization, compared to 67\% of the time at a static voltage. Moreover, we detect no iterations with two or more simultaneous outliers with voltage optimization, compared to 18\% of the time at $V_b = 0$.Finally, we compare the distributions of both average gate error (Fig. \ref{fig:4}\textbf{f}) and average $T_1$ (Fig. \ref{fig:4}\textbf{g}) for each of the 6 qubits over the entire experiment. We find a $36\%$ improvement in the average gate error with periodic local optimization ($\bar{\varepsilon}_P = 7.6 \times 10^{-4}$) compared to the static voltage $V_b = 0$ ($\bar{\varepsilon}_P = 11.9 \times 10^{-4}$). Similarly, we find a $17\%$ improvement in the average $T_1$ time with periodic local optimization ($\bar{T}_1 = 43.3$ \textmu s) compared to the static baseline voltage ($\bar{T}_1 = 36.9$ \textmu s).

These results demonstrate the advantage of using TIC-TAQ to locally optimize the noise environment across a multi-qubit device and confirm that the spectral diffusion of TLS defects is a dominant contributor to fluctuations in qubit performance. Notably, there was no recalibration of single qubit gate parameters over the entire 40 hour time period, highlighting the qubit's insensitivity to the TLS control parameter.

%% file: sections/M04_discussion.tex
In this work, we have introduced a scalable quantum control architecture, TIC-TAQ, to address the threat of TLS defects in a multi-qubit device. Our method enables local and independent control over each qubit's dissipative environment with no additional on-chip overhead, making it simple to integrate into current superconducting quantum processor designs. Leveraging in-situ spectral tuning of TLSs, we demonstrate improvements to qubit coherence times and single qubit gate performance that remain robust to changes in the TLS landscape. We also demonstrate the scalability of our method, showing that we can simultaneously stabilize single-qubit performance across 6 qubits over a 40 hour time period, with an overhead that does not scale with device size.

Our work represents a substantial step towards resolving spatial and temporal performance instabilities in superconducting quantum processors. Moreover, we expect TIC-TAQ to complement and enhance other strategies for addressing this challenge. For qubit frequency optimizers, the additional degree of freedom enabled by the spectral control of TLS defects significantly relaxes the constraints on viable operating frequencies. This should enable higher performance by opening up more possible configurations and allowing such optimizers to focus on other sources of error. Our technique will also accelerate efforts to reduce the density of TLS defects and better understand their origins. Sampling qubit coherence times over a range of applied voltages and frequencies provides a faster method for obtaining a holistic characterization of TLS loss, across a large-scale device. This should accelerate research on materials and fabrication processes with faster screening and process characterization of both test devices and full-scale processors. 
 
Given its effectiveness and simplicity of integration, we expect TIC-TAQ to become an important and widely adopted tool for maximizing the stability and performance of state-of-the-art superconducting quantum processors.

%% file: sections/S01_device.tex
\label{sec:device}
Our device consists of 16 fixed-frequency transmon qubits, each dispersively coupled to an independent $\lambda/2$ coplanar waveguide (CPW) readout resonator. The qubits and resonators are separated into two nominally identical groups (A and B), with each group sharing a multiplexed readout bus. Care was taken in the design to ensure that the qubit $T_1$ times are not limited by excess Purcell loss from either the readout circuitry or the TIQ-TAQ control line. However, the fabricated qubits have Josephson energies ($E_J$) that are $\sim$17-28 \% higher than intended, resulting in excess readout Purcell loss.

The $\lambda / 2$ readout resonators are coupled to the shared readout transmission line via an interdigitated capacitor (IDC) such that they form an ``intrinsic purcell filter" \cite{Sunada2022}. The open stub between the IDC and the end of the resonator is designed to be $\lambda/4$ at the qubit frequency, $f_q$, so that it forms an RF short at $f_q$. Similarly, the distance between each IDC and the open end of the readout feedline is also designed to be $\lambda / 4$ at $f_q$, creating an effective 2-pole bandstop filter between the qubit and the readout input.

The device is fabricated on a high-resistivity double-side polished silicon wafer ($\rho > 10$ k$\Omega$-cm) following a recipe based on Ref. \cite{Kreikebaum2020}. After cleaning the wafer with piranha at 120°C and hydrofluoric acid (HF) to remove organics and silicon oxide, a 200 nm layer of niobium is deposited with DC magnetron sputtering. The superconducting base layer, consisting of qubit capacitor pads, CPW resonators and control wiring, is defined by patterning a 1 \textmu m MiR701 photoresist with a direct-write maskless aligner. The exposed features are developed for 60 seconds in MF26A developer followed by ICP-RIE etching with BCl$_3$/Cl$_2$. The etch is monitored with end point detection, and we over-etch $\sim$70 nm into the silicon substrate to reduce the electric field participation of the qubit mode in the lossy silicon-air interface. The patterned wafer is cleaned in a high pressure N-methylpyrrolidone (NMP) jet at 80C.

The wafer is cleaned once again with buffered oxide etch (BOE) to remove oxides before fabricating the Josephson junctions. A resist bilayer is spun, where MicroChem MMA-EL 13 is used as the undercut layer and AllResist GmbH AR-P 6200.9 is used as the upper layer. The junctions are defined using e-beam lithography and developed in n-Amyl acetate (at 0°C) for the AR-P6200.9 resist and a 3:1 mixture of isopropyl alcohol (IPA) and deionized (DI) water (at 10°C) for the MMA-copolymer resist. The junction is deposited via a triple-angle e-beam evaporation of aluminum films following the Manhattan-style technique. The junctions are lifted off in acetone at 67°C, and a gentle oxygen plasma cleaning is applied to the devices afterward. The galvanic contact between the Josephson junctions and the qubit capacitor pads is formed using an argon ion-milling bandadge process \cite{Dunsworth2017}. The bandage is lifted-off with a high pressure NMP jet at 80C. Airbridges are then fabricated by first patterning 2 \textmu m MiR701 photoresist \cite{Chen2014}. The resist profile is smoothed out in a reflow process at 180°C to form the bottom bridge support, and 500 nm of aluminum is deposited at 45° to cover the bridge support. The airbridges are then finished by patterning of another layer of photoresist (MiR701, 1\textmu m) and wet etching the aluminum at 60°C. Finally, the fabricated wafer is diced. Chips are cleaned by soaking in NMP at 80°C, followed by successive rinses in DI water, acetone, and IPA. Finally, the selected chip is wirebonded to a printed circuit board and packaged in a copper box for measurement in a dilution refrigerator.

%% file: sections/S02_experimental-setup.tex
The packaged device is installed at the base stage of a Bluefors XLD1000 dilution refrigerator, with a base temperature of approximately 12 mK, for cryogenic characterization. The mounted assembly is enclosed in a copper IR shield coated with Sn/Pb solder on the outside to block and is further protected by an outer Mu-metal magnetic shield to suppress stray magnetic fields.

The device control wiring is illustrated in Fig. \ref{fig:circuit}. The qubit XY control and readout drive signals are directly synthesized using the FPGA-based QuBiC control platform \cite{Xu2021}. The resulting readout output signal is first amplified by a TWPA, whose pump tone is generated with a Holzworth HS9000B RF source, followed by a LNF HEMT amplifier at 4K. All microwave input lines (qubit XY, readout in, and TWPA) pass through attenuators thermally anchored at the 4K, still, and mixing chamber (MXC) stages, with additional low-pass and high-pass filters at the MXC to suppress input noise. The TLS control voltages are generated with a Yokogawa G200 voltage source, and routed through a phosphor-bronze wire loom with 12 twisted pair wires that are thermally anchored at the 4K and MXC stages. The DC voltage signal passes through a 10 kHz differential low-pass filter at the 4K stage, and is combined with the microwave qubit XY control signal using a Mini-Circuits ZX85-12G bias-tee mounted at the MXC stage, before being routed to the on-chip TIC-TAQ control line.

%% file: sections/S03_control-line.tex
\label{sec:cl-design}
In general, any capacitively coupled control line can be used for TLS manipulation by combining a DC voltage offset with the microwave RF qubit control pulses. However, most existing designs for qubit control focus on engineering sufficient coupling to the qubit without excess Purcell loss. This is typically realized by reducing the overall capacitance between the qubit capacitor pads and the control line, which inevitably limits the achievable TLS tuning sensitivities. The TIC-TAQ architecture circumvents this trade-off by exploiting symmetries in the qubit design to enable strong coupling to TLSs while still maintaining the necessary balance between sufficient coupling to the qubit mode and excess Purcell loss. We describe the design considerations for optimizing such a control line in the following sections.

\subsection{Circuit Analysis}
\label{sec:circuit-analysis}

\begin{figure}
    \centering
    \includegraphics[width=\columnwidth]{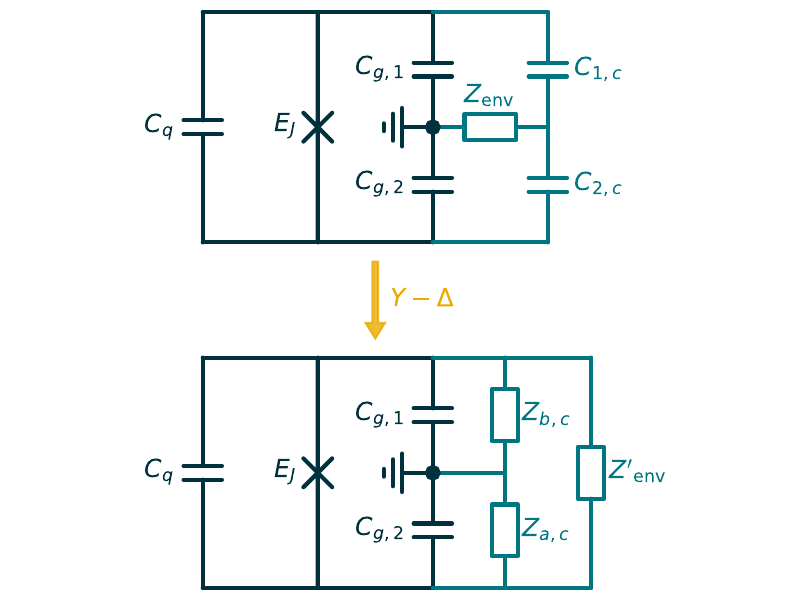}
    \label{fig:circuit}
    \caption{A circuit model for the reduced qubit-control line system showing the $Y-\Delta$ transform.}
\end{figure}

A lumped element circuit model of the qubit and control line is shown in Fig. \ref{fig:circuit}. The capacitance between the floating qubit pads is $C_{q}$. The qubit junction has a Josepshon energy $E_J$ with a corresponding inductance $L_J$. The qubit pads have capacitances to the control line $C_{1,c}$ and $C_{2,c}$, respectively. The control line (a semi-infinite CPW transmission line) is modeled as an effective impedance to ground,
\begin{equation}
    Z_{\mathrm{env}}=Z_{0}+\frac{1}{i\omega C_{g,c}}
\end{equation}
where $Z_{0}=50 \Omega$ is the characteristic impedance of the semi-infinite CPW transmission line and $C_{g,c}$ is the extra capacitance between the end of the control line and ground. Finally, the qubit pads have capacitances to ground $C_{g,1}$ and $C_{g,2}$, respectively. A $Y-\Delta$ transform can be applied to the control line sub-system (teal) to yield the equivalent circuit model in Fig. \ref{fig:circuit} with
\begin{align}
\begin{split}
    Z_{a,c}&=\frac{1+i\omega Z_{\mathrm{env}}(C_{1,c}+C_{2,c})}{i\omega C_{2,c}} \\
    Z_{b,c}&=\frac{1+i\omega Z_{\mathrm{env}}(C_{1,c}+C_{2,c})}{i\omega C_{1,c}} \\
    Z_{\mathrm{env}}'&=\frac{1+i\omega Z_{\mathrm{env}}(C_{1,c}+C_{2,c})}{-\omega^{2} Z_{\mathrm{env}C_{1,c}C_{2,c}}}.
\end{split}
\end{align}
Now we can use standard rules for simplification of parallel and series impedances/admittances to show that the qubit junction is shunted by an effective capacitance
\begin{equation}
    C_{\Sigma}=C_{q}-i\frac{d}{d\omega}\text{Im}\left(Y_{\mathrm{eff}}(\omega)\right)
\end{equation}
and effective resistance
\begin{equation}
    R=\frac{1}{\text{Re}\left(Y_{\mathrm{eff}}(\omega)\right)}.
\end{equation}
The qubit relaxation rate due to Purcell decay through the control is
\begin{equation}
    \Gamma=\frac{1}{RC_{\Sigma}}
\end{equation}
leading to a qubit quality factor 
\begin{equation}
    Q=\frac{\omega}{\Gamma}=\omega RC_{\Sigma}
\end{equation}
\cite{Esteve1986, Houck2008, Nigg2012}.

When the ground capacitances are equal, $C_{g,1}=C_{g,2}=C_{g}$, we find
\begin{equation}
    \Gamma=\frac{1}{C_{\Sigma}}\frac{\delta^{2}\omega^{2}C_{g}^{2}Z_{\mathrm{env}}}{\left(1+\frac{C_{g}}{C_{c}}\right)^{2}+4\omega^{2}Z_{\mathrm{env}}^{2}C_{g}^{2}}
\end{equation}
where
\begin{equation}
    \delta=\frac{C_{c,1}-C_{c,2}}{C_{c,1}+C_{c,2}}
\end{equation}
characterizes the asymmetry in the coupling of the qubit pads to the control line and $C_{c}=(C_{c,1}+C_{c,2})/2$ is the average coupling to the control line. Then $\delta=0$ ($C_{c,1}=C_{c,2}$) represents a symmetric control line design with no Purcell decay. Therefore, we can design large $C_{c,1}$ and $C_{c,2}$ to allow strong tuning of TLS in the vicinity of the qubit while independently maintaining $\delta \ll 1$ to limit Purcell decay.

More generally, we can view the circuit as a capacitive voltage divider that sets the voltages $V_{1}$ and $V_{2}$ on the qubit pads relative to an applied voltage on the control line. This is particularly useful when $C_{g,1}\neq C_{g,2}$. When the circuit is ``symmetric'' so that $V_{1}=V_{2}$ the differential qubit mode is not driven (and consequently does not decay) through the control line. From a design perspective we can start from a circuit that is symmetric in this way and introduce a small asymmetry to allow for qubit XY control. 

Finally, we note that the real chip design also included the effects due to capacitance between the qubit pads and the readout resonator. Analyzing the full circuit we find that the capacitance to the readout resonator simply renormalizes the ground capacitances, $C_{g,i} \rightarrow C_{g,i}'$, in a straightforward way. 

\subsection{EM Simulation}

\begin{figure}
    \centering
    \includegraphics[width=1\columnwidth]{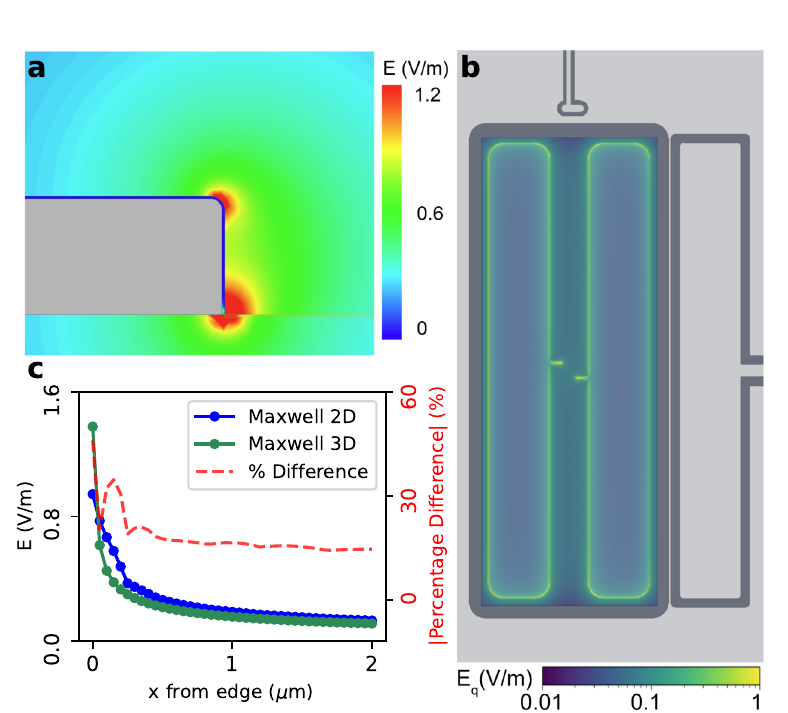}
    \caption{\textbf{Electromagnetic simulation of the qubit.} \textbf{a, b,} Simulation of the qubit plasma oscillation mode with $V_\text{rms} = 5.016$ \textmu V in Maxwell 2D and 3D, respectively. The minimum mesh size is 4 nm in 2D and 800nm in 3D. \textbf{c}, A comparison between the 2D and 3D simulations show agreement within a $\sim50\%$ difference, with the largest deviation happening between 4nm to 400nm, which the mesh in the 3D simulation does not cover.}
    \label{fig:M2D3D}
\end{figure}

To demonstrate the advantage of the IC-TAQ control line design, we analyze the strength of the TLS tuning field, $\mathbf{E}_{\text{TLS}}$, across several designs for capacitively coupled XY control lines. We perform finite element method (FEM) electromagnetic (EM) field simulations using Ansys Maxwell. In Fig. \ref{fig:M2D3D}\textbf{a}, we first show that the field of the qubit mode $\mathbf{E}_{q}$ ($V_{\text{rms}} \sim 5$\textmu V) is concentrated within a few hundred nanometers of the edges of the capacitor pads. Since the TLS-qubit coupling is proportional to $|\mathbf{E}_{q}|$, relevant TLS defects affecting the qubit primarily reside at these surfaces where the qubit electric field is sufficiently strong \cite{Lisenfeld2019, Bilmes2020}. In Fig. \ref{fig:M2D3D}\textbf{b}, we perform the same simulation with the Maxwell 3D solver. Generally, 3D solvers under-perform in resolving fields at the nanometer-scale compared to 2D solvers because they require more computational power. For example, the smallest mesh here is 4 nm in the 2D simulation and 800 nm in the 3D case. Still, the 3D simulation qualitatively captures the edge-concentrated distribution of $\mathbf{E}_{q}$. In Fig. \ref{fig:M2D3D}\textbf{c}, we compare the simulation results within 2 \textmu m of the edges of the capacitor pads and plot the relative deviation between the two solvers. We can linearly interpolate the 3D simulation results to get nm-scale information between adjacant mesh points. The resulting difference between the results from the 2D and 3D solvers is at most $\sim50\%$ across the region of interest. As expected, the maximum deviation happens over the intermediate regime between the smallest node distance in the 2D mesh and that in the 3D mesh (2 nm $<$ x $<$ 400 nm). A similar deviation is expected in the case of $\mathbf{E}_{TLS}$. As such, we will utilize the 3D solver with a similar mesh size to capture the geometric effects, but with the understanding that such a simulation has a limited spatial resolution in this intermediate regime.

\begin{figure}[b]
    \centering
    \includegraphics[width=\columnwidth]{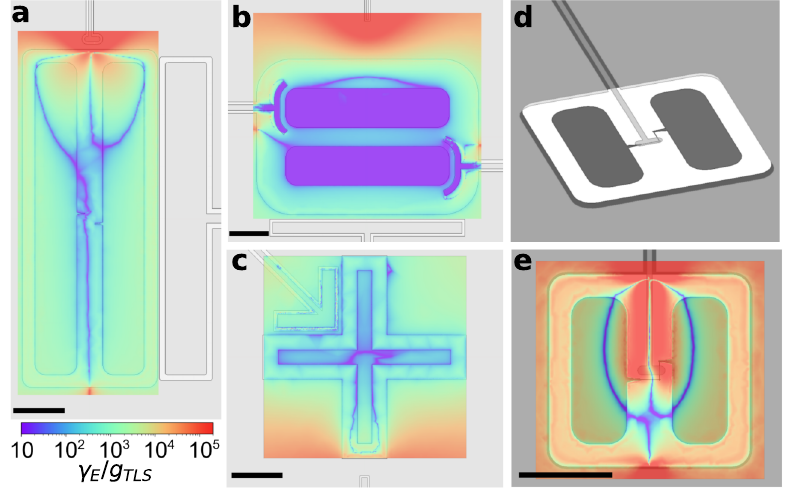}
    \caption{\textbf{Normalized tuning strength of qubit control-line designs.} Finite element simulations of $\gamma_{E}/g_{TLS}$ for: \textbf{a,} the TIC-TAQ control line used in this work, \textbf{b,} a representative floating qubit design \cite{Nguyen2024}, and \textbf{c,} a representative Xmon qubit design \cite{Sung2021}. \textbf{d, e,} A proposed implementation of the TIC-TAQ design in flip-chip architecture and the simulated normalized tuning strength. The black scale bars in each panel indicate a distance of $100$ \textmu m.}
    \label{fig:sim}
\end{figure}

In Fig. \ref{fig:sim}, we investigated three different qubit designs: \textbf{a} the TIC-TAQ control line design in this work, \textbf{b} a common floating transmon design \cite{Nguyen2024}, and \textbf{c} a ``Xmon" design \cite{Barends2013} from Ref. \cite{Sung2021}. To compare these designs, we define the normalized TLS tuning strength as the ratio $\gamma_{E}/g_\text{TLS}$, where
\begin{equation}
    \gamma_E = 2 \mathbf{p} \cdot \mathbf{E_{TLS}}/V_{DC},
\end{equation}
is the TLS tuning sensitivity to the applied electric field, and
\begin{equation}
    g_\text{TLS} =  \mathbf{p} \cdot \mathbf{E_{q}},
\end{equation}
is the coupling strength between the qubit and the TLS. For simplicity we choose $\mathbf{p} \parallel \mathbf{E_q}$. Intuitively, this ratio represents the degree to which the qubit can be decoupled from TLS defects, since $\gamma_E$ determines the achievable frequency shift of a TLS and $g_\text{TLS}$ sets the frequency scale at which a detuned TLS remains a dominant source of decoherence for the qubit. In practice, the detuning required to completely mitigate the detrimental effects of a TLS also depends on other properties, such as the coherence time of the defect.  Our simulation results highlight the advantage of the TIC-TAQ design. First, the TLS tuning strength is mostly $> 100$ V$^{-1}$ around the edges of the shunt capacitor pads, reaching values $>1000$ V$^{-1}$ close to the XY control line. In stark contrast, the floating transmon design example here shows a marginal normalized TLS tuning strength ($< 10$ V$^{-1}$) around the edges of the qubit pads. Since the control line approaches the qubit asymmetrically ($C_{c,1} \gg C_{c,2}$), the TLS tuning strength and the qubit mode coupling strength both scale with the qubit-control line separation. Maintaining a low Purcell loss rate consequently leads to a weak normalized tuning strength. Likewise with the Xmon design, there is no symmetry that can be exploited to independently set the control line coupling to the qubit mode relative to the TLS tuning strength. This design shows a significant tuning strength at one end of the qubit capacitor pad while leaving other areas only slightly tunable ($<40$ V$^{-1}$). In summary, our EM simulations show that the TIC-TAQ control design significantly increases both the strength and coverage of the TLS tuning field.

Implementing the TIC-TAQ design in a 3D-integrated (flip-chip) quantum circuit platform \cite{Foxen2018, Rosenberg2017} is also straightforward. As shown in Fig. \ref{fig:sim}\textbf{d}, the flip-chip geometry allows us to extend the end of the control line between the qubit capacitor pads (along the axis of symmetry) so that it ends directly above the junction. We note here that both the control line-induced Purcell loss and the total qubit capacitance are chosen to be as close as possible to the TIC-TAQ design used in this work (Fig. \ref{fig:sim}\textbf{a}). The smaller qubit footprint results from the presence of the additional silicon ($\varepsilon_r=11.7$) located above the qubit. Fig. \ref{fig:sim}\textbf{e} shows the normalized tuning strength where, compared to the planar designs, the normalized TLS tuning strength is two orders of magnitude stronger and covers nearly the entire area over which TLS defects are expected to couple to the qubit. This confirms the compatibility of the TIC-TAQ architecture with 3D-integration, which is necessary for building large-scale quantum processors.

%% file: sections/S04_TLS-fitting.tex
\label{sec:tls-fitting}

As discussed in the main text, the TLS spectrum is unique for each qubit. Here, we fit the voltage response of the TLS to the STM (Eq. \ref{eq:stm}), based on the example TLS tuning spectrum shown in Fig. \ref{fig:2}\textbf{a}. The extracted STM values can provide microscopic insights into each TLS that is coupled to the qubit. As discussed in Sec. \ref{sec:tuning}, AC Stark shift spectroscopy allows us to identify individual TLS defects from dips in the $P_1$ spectrum. After identifying peaks in each constant voltage slice, we select points belonging to a continuous tuning curve for fitting to the hyperbolic model. For this particular dataset, we were able to identify 11 unique TLS tuning curves that we could fit with reasonable fidelity.

In Fig. \ref{fig:tls-fit}\textbf{a}, we show the identified TLS tuning trajectories with their fits to the STM indicated by the black dotted lines. The extracted tuning strengths, $\gamma_E$, and tunneling energies, $\Delta_0$, are plotted in Fig. \ref{fig:tls-fit}\textbf{b}, with error bars indicating the fit error. As expected, the tunneling energies are randomly scattered around the qubit frequency. We note that the limited spectral range achievable with the AC Stark effect makes it difficult to obtain accurate fits when tunneling energy is far below the qubit frequency. However, the detection range can be extended by leveraging the higher transitions of the transmon. We find that the tuning strengths for the identified TLS defects vary by more than an order of magnitude, which can result from differences in their locations, the magnitudes of their electric dipole moments, or the directions of their electric dipole moments relative to the applied electric field.

%% file: sections/S05_crosstalk.tex
\label{sec:crosstalk}

The successful simultaneous mitigation of TLS-induced noise in a multi-qubit processor requires independent and local control between each control line and its target qubit's TLS environment. Unwanted DC crosstalk complicates the optimization process and increases the overhead of finding a simultaneous set of optimal voltage biases across number of qubits. We characterize the DC crosstalk in our system by simultaneously monitoring the TLS spectra of all 12 qubits with an AC Stark shift spectroscopy measurement, while sweeping the applied bias voltage on a single control line at a time. The results of these crosstalk measurements are shown in Fig. \ref{fig:crosstalk}. Of the 132 potential qubit-control line pairs that correspond to unintended crosstalk, we see a detectable level of crosstalk on 7 such pairs (red border). By cross-referencing this data with the physical layout of the device (Fig. \ref{fig:chip}), we find that crosstalk primarily occurs when a control line passes by a nearby qubit. We attribute this to the fact that while the control line is shielded from above by grounding airbridges, there is no shielding from below in the silicon substrate. Consequently, we expect the straightforward integration of superconducting through-silicon vias (TSV's) for shielding each control line to substantially reduce the level of crosstalk \cite{Hazard2023}.

%% file: sections/S06_optimization.tex
\label{sec:optimization}

Here we discuss the optimization strategy used for TIC-TAQ in greater detail. The primary goal of the optimization procedure is to determine the best operating voltage to apply with the TIC-TAQ control line at a given point in time $t_i$, where $i$ specifies the $i$-th iteration of the procedure.

For each iteration, we perform AC Stark shift spectroscopy on the qubit to characterize the TLS distribution in the spectral neighborhood of the qubit and its response to the voltage bias. We prepare the qubit in the $\ket{1}$ state, and measure the $\ket{1}$-state population $P_{1,i}(f, V)$ (after a fixed time delay) as a function of both the Stark-shifted frequency $f$ and the voltage bias $V$. We then average the traces over all frequencies as,
\begin{equation}
    \bar{P}_{1, i}(V) = \frac{1}{K}\sum_k P_{1,i}(f_k, V),
\end{equation}
where $K$ is the total number of sampled frequencies. Next, we take into account historical data by performing a weighted average over prior iterations:
\begin{equation}
    \mathcal{P}_{1,i}(V) = \frac{1}{N}\sum_{j=0}^i W(t_i, t_j)\bar{P}_{1,j}(V),
\end{equation}
where the normalization factor $\mathcal{N}$ is given by $N = \sum_{j=0}^i W(t_i, t_j)$. Finally, we select the optimal voltage for iteration $i$, $V^*_i$, as 
\begin{equation}
    V^*_i = \argmax_V \mathcal{P}_{1,i}(V).
\end{equation}
This can then be refined with a finer voltage scan around $V_i^*$, as discussed in Sec. \ref{sec:stability}.

The choice of voltages $V$ and frequencies $f$ over which to perform the stark shift spectroscopy scan depends on the precise details of the TLSs that couple to the qubit and in particular the distribution of tuning sensitivies, $\gamma_E$. For small $\gamma_E$, it is necessary to scan a larger voltage range to ensure that any TLS near the qubit transition frequency, $f_{01}$, can be tuned sufficiently far away from the transition. For large $\gamma_E$, it is necessary to perform a finer scan over voltages to ensure that no TLSs come into resonance at intermediate voltages. This, in principle, requires scanning over a large number of voltage points to accurately estimate the qubit's $T_1$ dependence on the bias voltage. However, we can reduce the number of voltage points by measuring the TLS distribution in a small neighborhood $\Delta f$ around $f_{01}$ to detect TLS at nearby frequencies for each voltage bias. Since the TLS frequency is approximately linear in $V$ for a sufficiently small voltage spacing, this has the same effect as performing a finer voltage scan. Thus, the trade-off between scanning over frequencies and voltages is primarily dependent on the wall-clock time of each measurement sweep.

In our experimental setup, the voltage bias is applied with a ``slow" DC source, while the AC stark shift tone is applied using the same ``fast" AWG that implements the qubit control pulses. As a result, there is a significantly larger time overhead for each additional voltage point in the sweep compared to additional frequency points. For the experiments described in Fig. \ref{fig:3} and Fig. \ref{fig:4}, we scan over 51 evenly spaced voltages between -10 V and 10 V, and 9 evenly spaced frequencies between $f_{01} \pm \Delta f$ where $\Delta f = 5$ MHz.

The choice of weighting function $W(t, t')$ depends primarily on the timescale of TLS spectral diffusion. In the limit where $W(t,t') = \delta(t - t')$ is the Dirac delta-function, the procedure does not take into account any historical data and may be sensitive to small fluctuations of TLS spectrum. However, this allows the optimization algorithm to adapt more quickly to an abrupt but substantive change to the TLS distribution. In contrast, stronger weighting of prior traces can promote greater stability in the choice of $V^*_i$ by prioritizing voltage biases with consistently higher $P_1$ values, but may be slower to adapt to changes. For the experiments described in Fig. \ref{fig:3} and Fig. \ref{fig:4}, we use a linear weighted average, 
\begin{equation}
    W(t_i, t) = \max\left(1 - \frac{t_i-t}{\tau}, 0\right)
\end{equation}
where the time constant $\tau$ was chosen to be 75 (60) minutes in Fig. \ref{fig:3} (Fig. \ref{fig:4}).

While the hyperparameters used in this work were found to work well, the optimal choice of hyperparameters for a given system has not been systematically studied and is the subject of ongoing research. This includes alternative weighting functions, which may be analytical (exponential weights) or empirically derived using machine learning techniques. In general, we can also incorporate more input data into the optimization procedure for different applications. For example, we can perform Stark shift spectroscopy on the higher levels of the transmon by preparing the transmon in the $\ket{s}$ state and measuring the population $P_s(f, V)$ as a function of $V$ and the stark shifted transition frequency $f_{s,s-1}$. Including data from both $s=1$ and $s=2$ would promote voltage biases $V^*_i$ that simultaneously improve both the coherence of the $\ket{1}$ and $\ket{2}$ states of the transmon, which can improve the performance of qutrit computation \cite{Goss2022} or two-qubit gate schemes like the diabatic CZ gate that temporarily populate the $\ket{2}$ state \cite{Barends2019, Sung2021}. It may also be advantageous to use RB as a cost function \cite{Kelly2014}, since this can capture additional detrimental effects induced by TLSs, such as frequency shifts or excess dephasing.